\documentclass[11pt]{article}
\usepackage{graphicx}
\usepackage{array}

 \newcommand{\btodk}{$B^{-}\ra D^0 K^{-}$}
 \newcommand{\btodkcp}{$B^{\pm}\ra D^0_{CP} K^{\pm}$}
 \newcommand{\btodp}{$B^{-}\ra D^0 \pi^{-}$}
 \newcommand{\btodh}{$B^{-}\ra D^0 h^{-}$}

 \newcommand{\dotokp}{$D^0\ra K^-\pi^+$}
 \newcommand{\dotokppp}{$D^0\ra K^-\pi^+\pi^+\pi^-$}
 \newcommand{\dotokppo}{$D^0\ra K^-\pi^+\pi^0$}
 \newcommand{\dotokk}{$D^0\ra K^-K^+$}

 \newcommand{\deltaemk}{$\Delta E_K$}
 \newcommand{\deltaemp}{$\Delta E_{\pi}$}

\newcommand{\BABARPubYear}    {02}

\newcommand{\BABARConfNumber} {018}
\newcommand{\SLACPubNumber} {9311}

\input pubboard/babarsym

\long\def\inst#1{\par\nobreak\kern 4pt\nobreak
    {\it #1}\par\vskip 10pt plus 3pt minus 3pt}

\begin{document}
{\pagestyle{empty}

\begin{flushright}
\babar-CONF-\BABARPubYear/\BABARConfNumber \\
SLAC-PUB-\SLACPubNumber \\
%hep-ex/\LANLNumber \\
July 2002 \\
\end{flushright}

\par\vskip 5cm

\begin{center}
\Large \bf Measurement of Branching Ratios and \CP\ Asymmetries in $B^-\ra D^0_{(\CP)}K^-$ Decays
\end{center}
\bigskip

\begin{center}
\large The \babar\ Collaboration\\
\mbox{ }\\
July 25, 2002
\end{center}
\bigskip \bigskip

\begin{center}
\large \bf Abstract
\end{center}

We present preliminary results of the analysis of $\B \to \Dz h$
decays, with $ h$ = $\pi, K$ and the $D^0$ reconstructed
in the channels $K^-\pi^+$, $K^-\pi^+\pi^+\pi^-$, 
$K^-\pi^+\pi^0$ and in the \CP\ eigenstate $K^-K^+$, using 
data collected by the \babar\ detector during the years 2000-2002
at the \pep2\ asymmetric-energy $B$ Factory at SLAC.   
We have measured the ratio of the branching fractions
\begin{eqnarray}
R\equiv\frac{\BR(B^-\ra D^0 K^-)}{\BR(B^-\ra D^0
\pi^-)}=(8.31\pm 0.35\pm 0.20)\%\ \nonumber
\end{eqnarray}
and the direct \CP\ asymmetry
\begin{eqnarray}
A_{\CP}\equiv\frac{\BR(B^-\ra D^0_{\CP} K^-)-\BR(B^+\ra D^0_{\CP} K^+)}{\BR(B^-\ra D^0_{\CP} K^-)+\BR(B^+\ra D^0_{\CP} K^+)}= 0.17\pm 0.23\, ^{+0.09}_{-0.07}\ .\nonumber
\end{eqnarray}

\vfill
\begin{center}
Contributed to the 31$^{st}$ International Conference on High Energy Physics,\\ 
7/24---7/31/2002, Amsterdam, The Netherlands
\end{center}

\vspace{1.0cm}
\begin{center}
{\em Stanford Linear Accelerator Center, Stanford University, 
Stanford, CA 94309} \\ \vspace{0.1cm}\hrule\vspace{0.1cm}
Work supported in part by Department of Energy contract DE-AC03-76SF00515.
\end{center}

\newpage
}

\begin{center}
\small

The \babar\ Collaboration,
\bigskip

%% author list as of 05-Jul-2002 (556 authors)
B.~Aubert,
D.~Boutigny,
J.-M.~Gaillard,
A.~Hicheur,
Y.~Karyotakis,
J.~P.~Lees,
P.~Robbe,
V.~Tisserand,
A.~Zghiche
\inst{Laboratoire de Physique des Particules, F-74941 Annecy-le-Vieux, France }
A.~Palano,
A.~Pompili
\inst{Universit\`a di Bari, Dipartimento di Fisica and INFN, I-70126 Bari, Italy }
% ...

\end{center}\newpage

\section{Introduction}
\label{sec:Introduction}
During the last ten years there has been growing theoretical interest
in knowing the decay rates for the processes \btodk\
\footnote[1]{Charge conjugation is implied here and 
throughout this paper unless explicitly stated.} and \btodkcp,
where $D^0_{\CP}$ indicates the \CP-even or \CP-odd states
$(D^0\pm \Dzb)/\sqrt{2}$. These modes are key ingredients for some of
the recently proposed methods for extracting the angle $\gamma$ of the
Cabibbo-Kobayashi-Maskawa quark mixing matrix in a theoretically clean
way~\cite{gronau1991,atwood1997}. 

In this paper we present an analysis of the decay \btodk\ in which the
$D^0$ is reconstructed in the non-\CP\ eigenstates $K^-\pi^+$,
$K^-\pi^+\pi^+\pi^-$, $K^-\pi^+\pi^0$,
% (which in the following sections
%will be sometimes referred to as ``non-\CP\ modes'') 
or in the \CP-even eigenstate $K^-K^+$.
The ratio $R$ between the branching fractions of \btodk\ and \btodp
\begin{eqnarray}
R\equiv\frac{\BR(B^-\ra D^0 K^-)}{\BR(B^-\ra D^0
\pi^-)}
\end{eqnarray}
is measured, along with the ratio $R_{\CP}$ between the branching
fractions of $B^\pm\ra D^0_{\CP}\,K^\pm$ and $B^\pm\ra
D^0_{\CP}\,\pi^\pm$
\begin{equation}
R_{\CP}\equiv\frac{\BR(B^-\ra D^0_{\CP} K^-)+\BR(B^+\ra D^0_{\CP} K^+)}{\BR(B^-\ra D^0_{\CP} \pi^-)+\BR(B^+\ra D^0_{\CP} \pi^+)}.
\end{equation}
The yields of $B^-\ra D^0_{\CP}\,K^-$ and $B^+\ra D^0_{\CP}\,K^+$ are
separately extracted, and the \CP\ asymmetry 
\begin{equation}
A_{\CP}\equiv\frac{\BR(B^-\ra D^0_{\CP}\,K^-)-\BR(B^+\ra D^0_{\CP}\,K^+)}{\BR(B^-\ra
D^0_{\CP}\,K^-)+\BR(B^+\ra D^0_{\CP}\,K^+)}
\end{equation}
is measured.

The \btodk\ decay was first observed by the CLEO
collaboration~\cite{cleo_btdk}. Using a sample of 3.1 \invfb
collected at the \FourS\ resonance, CLEO measured $R=(5.5\pm1.4\pm0.5)\%$.
Recently, based on $10.4$~\invfb\ collected at the \FourS, Belle has measured
$R=(7.9\pm0.9\pm0.6)\%$~\cite{belle_btdk}. Belle also reported, using
a sample of $29.1$~\invfb, measurements of the direct \CP\ asymmetries
for the \CP-even and \CP-odd modes, $A_{\CP}(+1)=(0.29\pm0.26\pm0.05)\%$ and
$A_{\CP}(-1)=(-0.22\pm0.24\pm0.04)\%$~\cite{belle_bcptdk}.

\section{The \babar\ detector and dataset}
\label{sec:babar}
The data used in this analysis were collected with the \babar\ detector
at the \pep2\ storage ring during the years 2000-2002. 
The sample corresponds to an integrated luminosity of about 75 \invfb
accumulated %near 
at the \FourS\ resonance (``on-resonance'') and about 10
\invfb\ accumulated at a center-of-mass (CM) energy about 40 \mev
below the \FourS\ resonance (``off-resonance''). Data taken below the
\FourS\ are used for continuum background studies. 
The on-resonance sample corresponds to $(81.1 \pm 0.9)\times 10^6$ 
\BB\ pairs. 
The results regarding the non-\CP\ modes are based on a subsample of
data collected in the years 2000-2001; this corresponds to an
integrated luminosity of about 56 \invfb\ accumulated %near 
at the \FourS\ resonance, which in turn corresponds to $(61.2 \pm
0.7)\times 10^6$ \BB\ pairs.

\pep2\ is an \epem\ storage ring operated with asymmetric beam energies, 
producing a boosted ($\beta\gamma = 0.55$) \FourS\ along the 
collision axis. 
\babar\ is a solenoidal detector optimized for the asymmetric beam
configuration at PEP-II and is described in detail in Ref.~\cite{ref:babar}.
Charged particle (track) momenta are measured in a tracking system
consisting of a 5-layer, double-sided, silicon vertex tracker (SVT) and a
40-layer drift chamber (DCH) filled with a gas mixture of helium and
isobutane, both operating within a $1.5\,{\rm T}$ superconducting
solenoidal magnet. Photon candidates are selected as local maxima of
deposited energy in an electromagnetic calorimeter (EMC) consisting
of 6580 CsI(Tl) crystals arranged in barrel and forward endcap
subdetectors.
Particle identification is performed by combining information from
ionization measurements ($dE/dx$) in the SVT and DCH, 
and the Cherenkov angle $\theta_C$ measured by a
detector of internally reflected Cherenkov light (DIRC). The DIRC
system is a unique type of Cherenkov detector that relies on total internal
reflection within the radiating volumes (quartz bars) to deliver the
Cherenkov light outside the tracking and magnetic volumes, where the
Cherenkov ring is imaged by an array of $\sim$11000 photomultiplier tubes.

\section{Event selection}
\label{sec:Analysis}
We reconstruct $B$ mesons decaying to a \Dz\ meson and a charged
prompt track $h$, where $h$ is a pion or a kaon. $D^0$ meson candidates are
reconstructed in four decay modes: 
\dotokp, \dotokppp, \dotokppo\ and \dotokk. The decay \btodp\ is used
to evaluate detector resolutions, systematic uncertainties and to
normalize branching fractions.

All charged tracks are reconstructed in the drift chamber and/or the vertex
detector, and their parameters determined with the pion mass hypothesis. 
In order to reduce the combinatorial background, only charged
tracks with momentum greater than 150 MeV/c are used in the
reconstruction of \dotokppp\ and \dotokppo; the prompt track $h$ is
required to have momentum greater than 1.4 \gevc.
Particle ID information from the drift chamber ($dE/dx$) and, when
available, from the DIRC is required to be consistent with the kaon
hypothesis for the $K$ meson candidate from the \Dz. 
In order for the prompt track $h$ to be identified as a pion or a kaon
we require that its Cherenkov angle be reconstructed in the DIRC with
at least five photons. We reject candidate tracks
whose Cherenkov angle is within 3 standard deviations ($\sigma$, where
$\sigma$ means the experimental resolution) from the
expected angle for the proton hypothesis, and candidate tracks that are
identified as electrons by the DCH and the EMC.

Candidate \piz\ mesons are reconstructed from a combination of two photon
candidates. Photon candidates are selected as showers in the EMC that have the
expected lateral shape, are not matched with any charged track, and
have a minimum energy of 70 \mev. The $\gamma\gamma$ invariant mass is required
to be in the range 124--144 \mevcc, and the total energy of the $\gamma$
pair must exceed
200 \mev. The $\piz\ra\gamma\gamma$ candidates are then kinematically
fit with their mass constrained to the nominal \piz\ mass~\cite{ref:PDG}.

The invariant mass of \dotokp, \dotokppp\ and \dotokk\ candidates is required
to be within 3 $\sigma$ of the mean mass value. As the combinatorial
background of the \dotokppo\ decays is larger due to the presence of
a neutral pion, a 2-$\sigma$ cut on the invariant mass is applied for
this mode. \Dz\ candidates are then kinematically
fit with their mass constrained to the nominal \Dz
mass~\cite{ref:PDG}, in order to improve momentum determination.

We reconstruct $B$ meson candidates by combining a \Dz\ candidate
with a track $h$. For the non-\CP\ modes, the charge of the bachelor
track $h$ must match that of the kaon from the $D^0$ meson decay.
We exploit our knowledge of the initial state momentum and energy to
select $B$ meson candidates by defining two largely uncorrelated variables.
The first is the \emph{beam-energy substituted mass} $\mes = \sqrt{E^2_{\rm
b}-\mathbf{p}_B^2}$, where $E_{\rm b} =(s/2 + \mathbf{p}_i
\cdot\mathbf{p}_B)/E_i$, $\sqrt{s}$ and $E_i$ are the total energies
of the \epem\ system in the CM and lab frames, respectively, and
$\mathbf{p}_i$ and $\mathbf{p}_B$ are the momentum vectors in the lab
frame of the \epem\ system and the $B$ candidate, respectively. 
The second is the \emph{energy difference}
$\Delta E$, which is defined as the difference between the energy of the
$B$ candidate and half the energy of the \epem\ system, computed in
the CM system. The \mes\ resolution is dominated by the beam energy
spread, while for $\Delta E$ the main contribution comes from the
measurement of particle energies in the detector. 

The \mes\ distribution for \btodh\ signals does not depend on whether the
prompt track $h$ is a pion or a kaon, nor, to first order, on the \Dz
momentum resolution. 
This has been checked both for signal Monte Carlo and for data. The
\mes\ resolution for the \btodh\ channels is found to be $2.6 \mevcc$, 
regardless of the \Dz\ decay mode. 

In contrast, the \DeltaE\ distributions depend on the nature of 
the bachelor track $h$ and on the resolution for the \Dz\ meson
momentum. We evaluate $\Delta E$ with the kaon
mass hypothesis (and denote it by \deltaemk\ to avoid confusion) in such a
way that the distributions are centered near 0
for \btodk\ events and shifted by approximately $42 \mev$ for \btodp\ events.
The \deltaemk\ resolution is typically $17\mev$ for all \Dz\ decay modes.

$B$ candidates are selected in the range $5.2<\mes<5.3\gevcc$ and
$-0.100<\Delta E<0.130\gev$. For events with multiple $B$ candidates,
the best candidate is chosen based on the values of the $D^0$
invariant mass and \mes. 

\section{Background rejection}\label{sec:bkg}
The physics backgrounds for the considered modes originate both from 
the continuum production of light quarks, $e^+e^-\ra
q\overline{q}\ (q=u, d, s, c)$, and from \BB\ events. 
In the center-of-mass frame, the continuum
background typically exhibits a two-jet structure; in contrast,
the low momentum of the $B$ meson in the decay $\FourS \ra \BB$ leads
to a more spherically symmetric event.  We exploit this topology
difference between signal and continuum background by making use of two
event-shape quantities.

The first variable is the normalized second Fox-Wolfram
moment~\cite{ref:fox_wol}, $R_2\equiv\frac{H_2}{H_0}$,
where $H_l$ is the $l$--order Fox-Wolfram moment. $R_2$ is required to
be less than 0.5 for all the selected events. 

The second quantity is the angle $\theta_T$ between the thrust axes,
evaluated in the center-of-mass frame, of the $B$ candidate and the remaining
charged and neutral particles in the event. The absolute value of the
cosine of this angle is strongly peaked near 1 for continuum events
and is approximately uniform for \BB\ events. $|\cos\theta_T|$ is
required to be less than 0.9 for the \dotokp\ mode, and this value
is tightened to 0.7 for \dotokppp\ and \dotokppo\ modes, which suffer
from larger combinatorial background.

For the \dotokk\ mode an additional quantity, the $D^0$ rest frame decay angle
$\theta_{KK}$, is used in conjunction with  $|\cos\theta_T|$. 
The angle $\theta_{KK}$ is defined as the angle between the direction of the
$D^0$ calculated in the rest frame of the $B$ and the direction of one
of the decay products of the $D^0$ calculated in the rest frame of the
$D^0$. The distribution of $\cos\theta_{KK}$ is flat for signal and peaked at
$\pm 1$ for fake $D^0$ background.
$|\cos\theta_T|$ and $\cos\theta_{KK}$ are uncorrelated for signal
but \emph{not} for continuum background. This correlation is
exploited to make a more efficient cut in the
$\cos\theta_T-\cos\theta_{KK}$ plane. 

The main contributions from \BB\ background come
from the processes 
 $B^-\ra D^{*0}h^-$ ($h=\pi\,,K$), $B^-\ra D^0\rho^-$ and from
mis-reconstructed \btodh\ decays.  
Background from the charmless three-body process $B^-\ra K^-K^+K^-$
is potentially the most critical for the \btodk, \dotokk\ decay,
because it consists of three kaons coming from a $B$ meson
and hence is characterized by the same \DeltaE\ and \mes
distribution as the signal. Results from recent
studies~\cite{ref:belle_btohhh} of the resonant composition of such a
decay have been used to estimate this non-negligible background. 

The total number of $B$ candidates and the final selection efficiency
$\varepsilon$ for each mode are summarized in Table~\ref{tab:eff}.
The small differences between \btodp\ and \btodk\ efficiences
arise both from reconstruction (kaons have lower $dE/dx$ than pions
and hence fewer hits associated to the track) and from candidate
selection (the rejection of electrons and protons has different
efficiencies for kaons and pions).

\begin{table}[!htb]
\caption{Number of candidates selected for the maximum-likelihood fit
and final selection efficiencies.
% The candidates for the non-\CP\ modes
%have been selected on the 2000-2001 data sample; the candidates
%for the \dotokk\ mode have been selected on the full 2000-2002 data
%sample. 
The selection efficiency has been evaluated on simulated
events.}
\begin{center}
\begin{tabular}{l|c|c|c|c}
\hline
\hline
$D^0$ decay mode & events selected &$\varepsilon($\btodp$)$ & $\varepsilon($\btodk$)$
&$\int\mathcal{L}\,dt \ (\textrm{fb}^{-1})$\\ 
\hline
$K^-\pi^+$ & 12606 & $(43.65\pm 0.12)\%$ & $(42.17\pm 0.31)\%$ & 56.0\\
$K^-\pi^+\pi^+\pi^-$ & 9782& $(14.53\pm 0.17)\%$ & $(13.55\pm 0.22)\%$ & 56.0\\
$K^-\pi^+\pi^0$ & 8177& $(8.04\pm 0.11)\%$ & $(8.12\pm 0.18)\%$ & 56.0\\ 
$K^-K^+$  & 2389 & $(34.5\pm 0.2)\%$ & $(33.7\pm 0.3)\%$ & 74.8\\
\hline
\hline
\end{tabular}
\end{center}
\label{tab:eff}
\end{table}

\section{Signal extraction}
For each \Dz\ decay mode an extended unbinned maximum-likelihood fit
to the selected data events determines the signal and background yields $n_i$
($i=1$ to $M$, where $M$ is the total number of signal and background 
species). We consider two kinds of signal events, \btodp\ and \btodk,
and four kinds of background events: candidates selected either from
continuum or from \BB\ events, in which the prompt track $h$ is either
a pion or a kaon. In the case of \btodh, \dotokk\ events, the fit is
also performed on the $B^+$ and $B^-$ subsamples separately.

The input variables to the fit are \mes, \deltaemk, and a particle
identification 
probability for the bachelor track $h$ based
on the Cherenkov angle $\theta_C$, the momentum $p$ at the DIRC and
the polar angle $\theta$ of the track.
The extended likelihood function $\cal L$ is defined as
\begin{equation}
{\cal L}= \exp\left(-\sum_{i=1}^M n_i\right)\, \prod_{j=1}^N
\left[\sum_{i=1}^M n_i {\cal P}_i\left(\vec{x}_j;
\vec{\alpha}_i\right) \right]\, .
\end{equation}
The exponential factor in the likelihood accounts for Poisson fluctuations
in the total number of observed events $N$. 
The  $M$ functions ${\cal P}_i(\vec{x}_j;\vec{\alpha}_i)$ are the
probability density functions (PDFs) for the variables
$\vec{x}_j$, given the set of parameters $\vec{\alpha}_i$. They are
evaluated as a product $\mathcal{P}_i=\mathcal{P}_i(\DeltaE_K,m_{\rm
ES})\times\mathcal{P}_i({\rm DIRC})$,
since \mes\ and \deltaemk\ are not correlated with the Cherenkov angle
of the prompt track. 

The correlation between \mes\ and \deltaemk\ can be neglected for signal  
events and for the combinatorial background from the continuum:
$\mathcal{P}_i(\DeltaE_K,m_{\rm ES}) =
\mathcal{P}_i(\DeltaE_K)\times\mathcal{P}_i(m_{\rm ES})$. However it cannot
be neglected for the \BB\ background component, for which we use 
a two-dimensional PDF determined from simulated events through a method based 
on the \emph{Kernel Estimation}~\cite{ref:kernel1} technique. 

The parameters for the \deltaemk\ and \mes\ distributions for continuum
background are determined from off-resonance data. 
The background shape in
\deltaemk\ is parameterized as a linear polynomial, while the \mes
shape is parameterized by an ARGUS threshold function~\cite{ref:argus} 
$f(\mes)\propto \mes\sqrt{1-x^2}\exp[-\xi(1-x^2)]$, where $x=\mes/m_0$
and $m_0$ is the average CM beam energy. 

The \mes\ PDF for signal events is determined from a pure
sample of \btodp, \dotokp\ decays selected from on-resonance data by
applying the standard selection plus a 3 $\sigma$ cut on \deltaemp. 
\BB\ background is negligible in this channel; therefore
the \mes\ distribution is fit by a Gaussian signal over a
background described by an ARGUS function. 
The resulting Gaussian shape is used to parameterize the \mes\ PDF
for the \btodh\ signals. 

The parameterization of the \deltaemk\ distribution for \btodk
events is deduced from that of \deltaemp\ for \btodp\ events
in the data. The \deltaemp\ distribution for a pure sample of \btodp
decays, selected by applying the standard selection plus a 3 $\sigma$ cut on
\mes\ and the requirement that the bachelor track $h$ not be
consistent with the kaon hypothesis,
is fit by a Gaussian signal over a linear background. The
resulting Gaussian shape is used to parameterize the \deltaemk\ PDF
for the \btodk\ signal. We also use the \deltaemp\ resolution
together with the known momentum spectrum of the bachelor track in the
laboratory frame to generate a ``translated'' \deltaemk\ distribution
for \btodp\ events:
for each generated value of \deltaemp\ and $p$, we calculate the
energy shift according to  
$\Delta E_{\rm shift}\equiv \Delta E_K-\Delta
E_{\pi}=\gamma\left(\sqrt{m_K^2+\vec{p}^{\,2}}-\sqrt{m_{\pi}^2+\vec{p}^{\,2}}\right)$, where $\gamma$ is the Lorentz
boost of the center-of-mass frame.     
The resulting \deltaemk\ distribution is empirically parameterized by the 
sum of two Gaussians.

The $\theta_C$ PDFs are derived from kaon and pion tracks in the
momentum range of interest from a sample of $D^{*+}\to\Dz\pip$
($\Dz\to \Km\pip$) decays. Since the $\theta_C$
resolution depends in principle uniquely on the number of found
photons, and the number of expected photons depends on both the track
polar angle and momentum, we use this sample to
parameterize the $\theta_C$ resolution as a function of track polar
angle and momentum extrapolated to the DIRC.

\begin{figure}[!htb]
\begin{center}
\includegraphics[width=8cm]{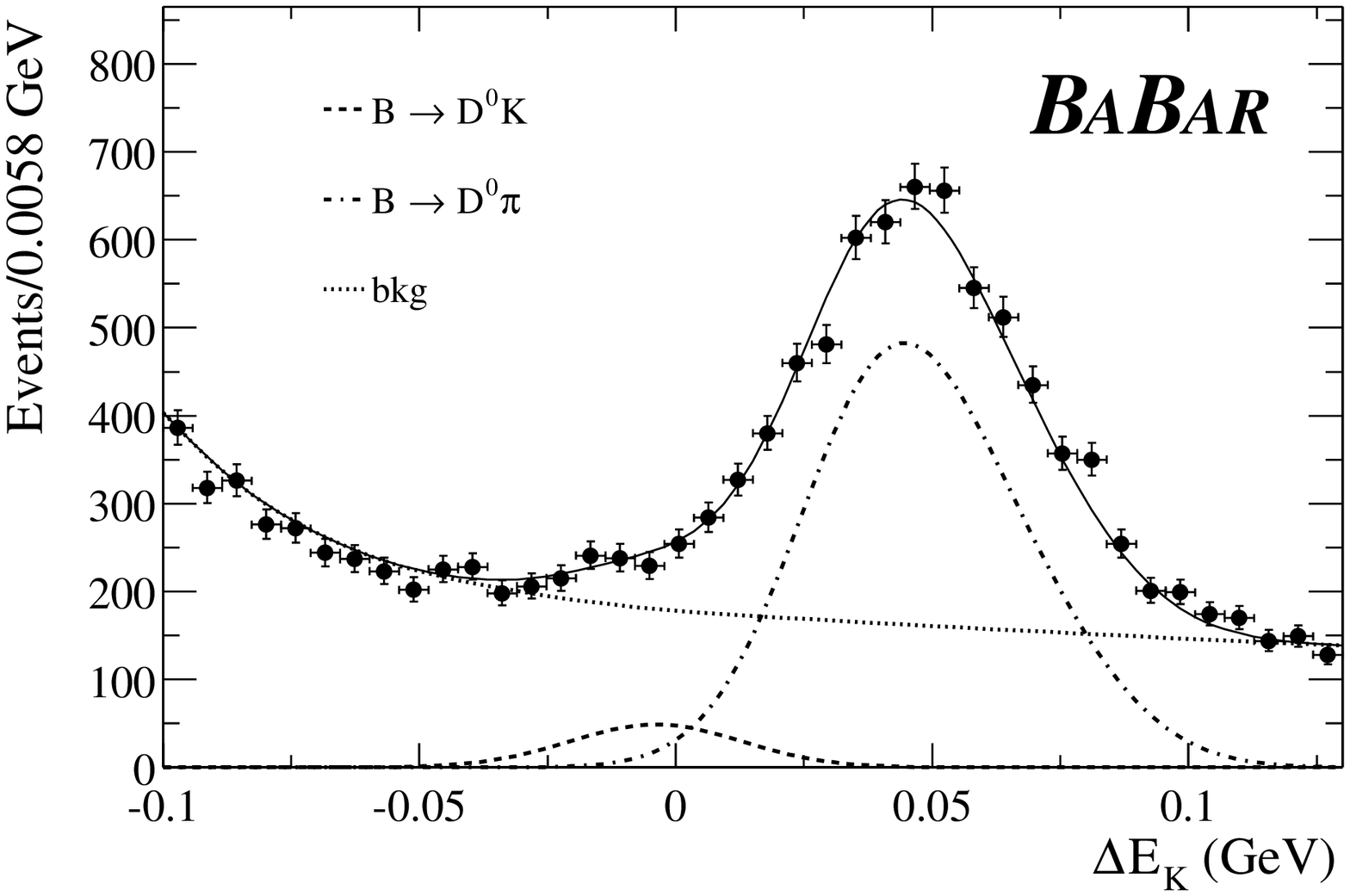}\includegraphics[width=8cm]{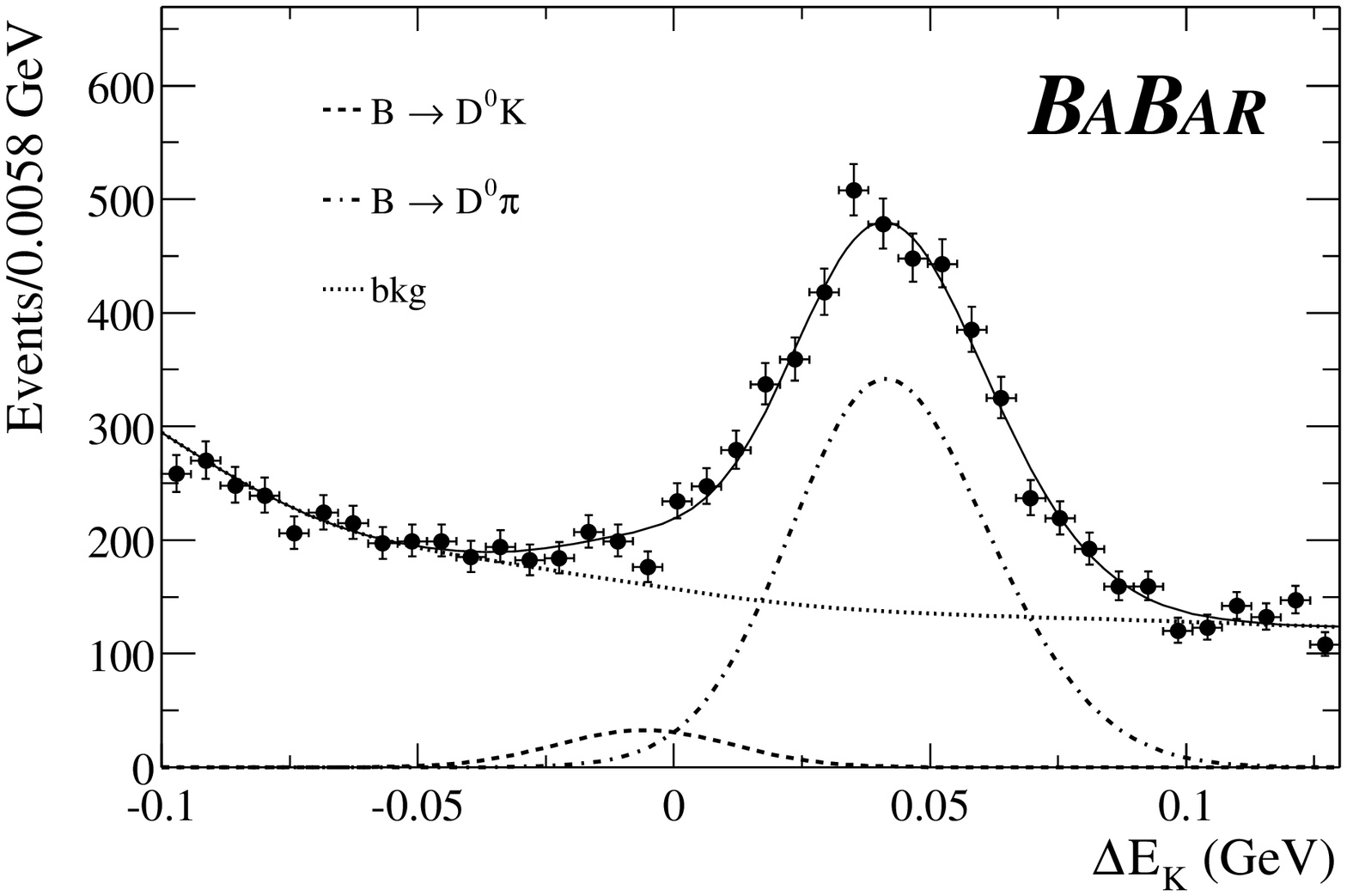}
\includegraphics[width=8cm]{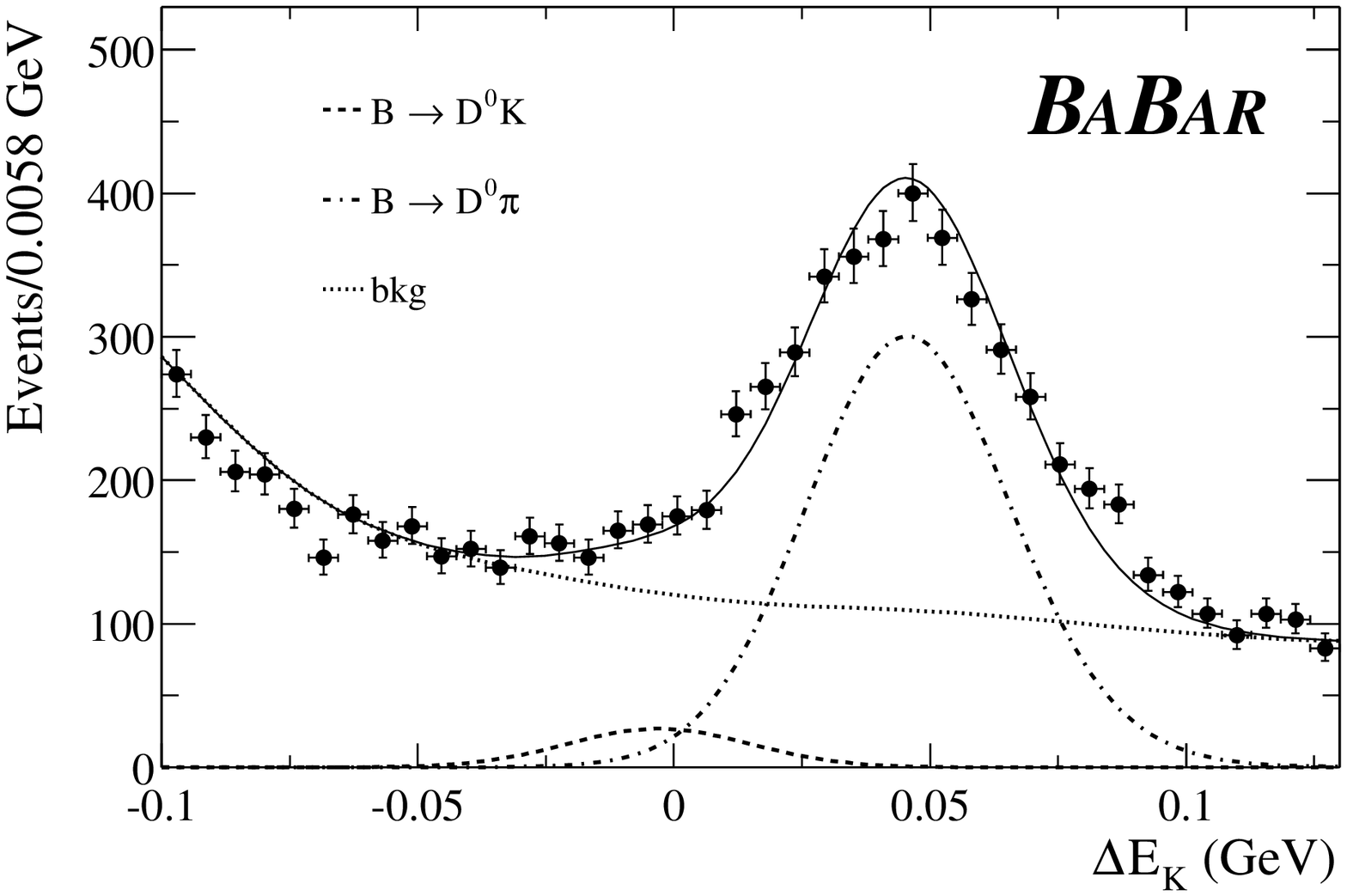}\includegraphics[width=8cm]{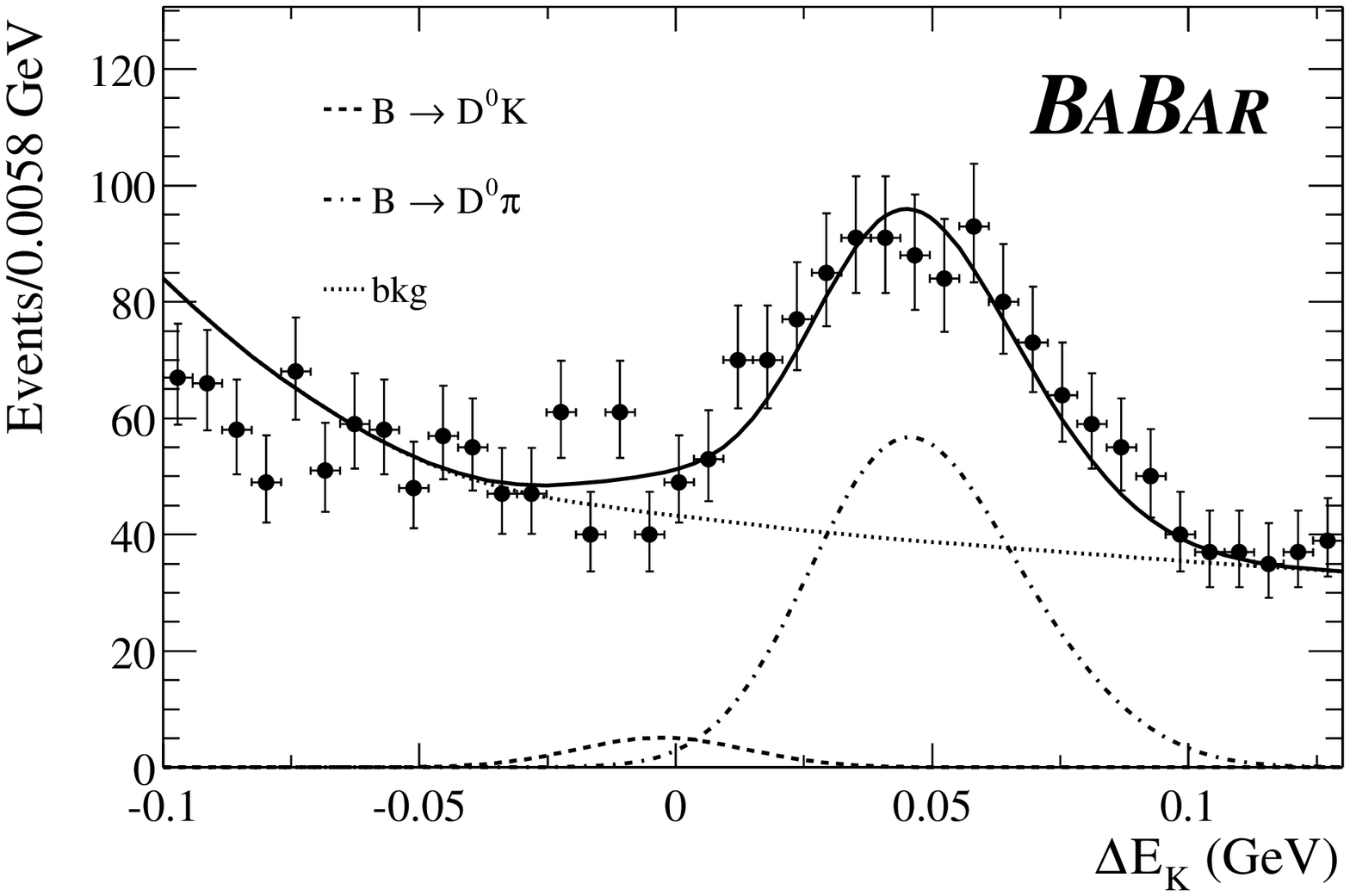}
\caption{\deltaemk\ distribution of the \btodh\ candidates selected in
the data sample. As described in the text, tracks $h$ consistent with
the electron or proton hypothesis have been explicitly removed. 
Top left: \dotokp; top right: \dotokppp;
bottom left: \dotokppo; bottom right: \dotokk. The solid curve
represents the projection on the \deltaemk\ axis of the resulting fit
probability density function, scaled by the number of candidates in
the sample. The contribution from the two signals and from the total
background is also shown.}
\label{fig:fit}
\end{center}
\end{figure}

The results of the fit are summarized in Table~\ref{tab:fitresults}
and the \deltaemk\ distributions are shown in Fig.~\ref{fig:fit} for
all the $D^0$ modes. 
The projection of the likelihood fit is overlaid on the
distributions. The individual contributions of each of the signals and
the total background are shown.
 
\begin{figure}[!htb]
\begin{center}
\includegraphics[width=8cm]{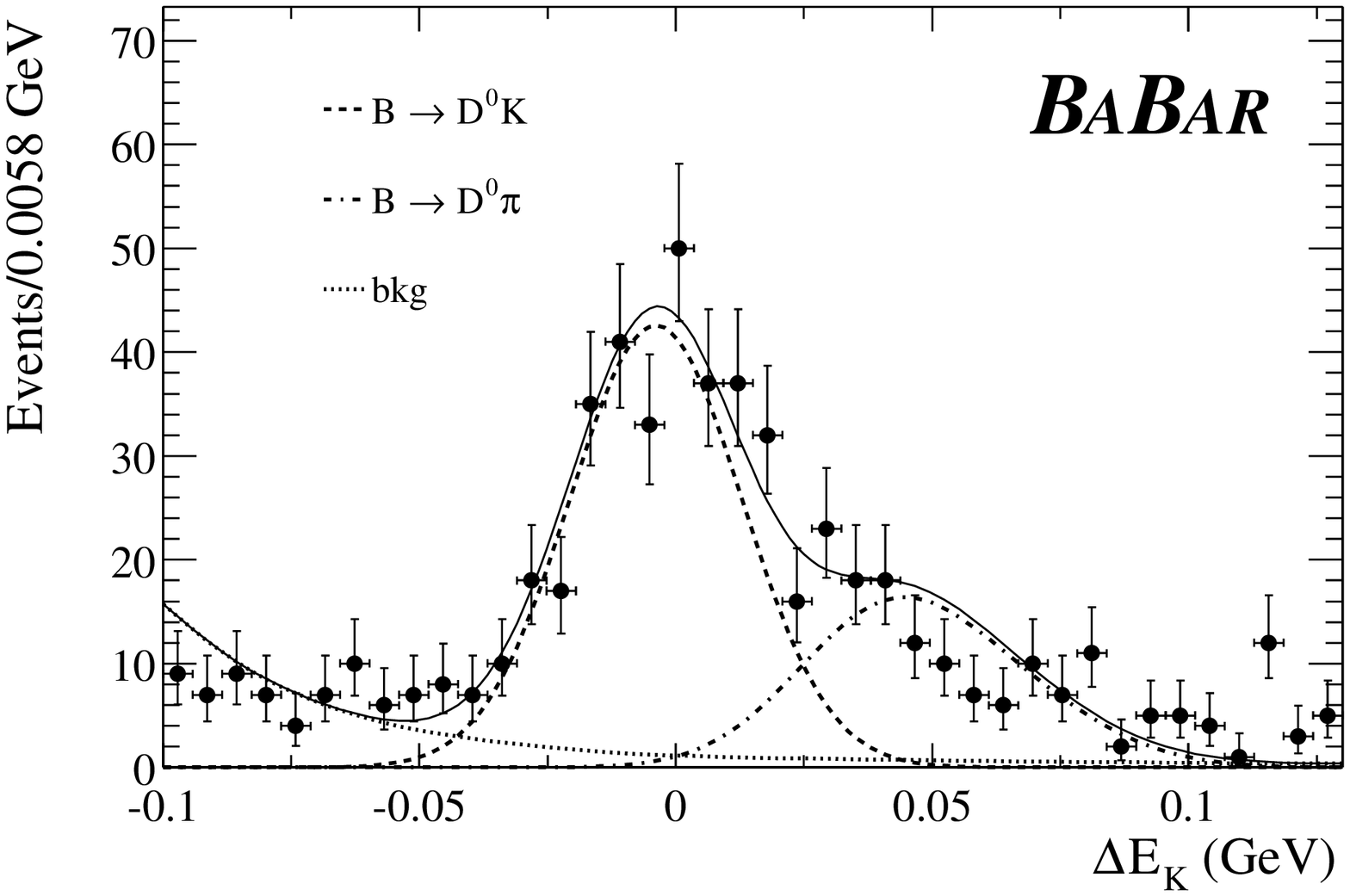}\includegraphics[width=8cm]{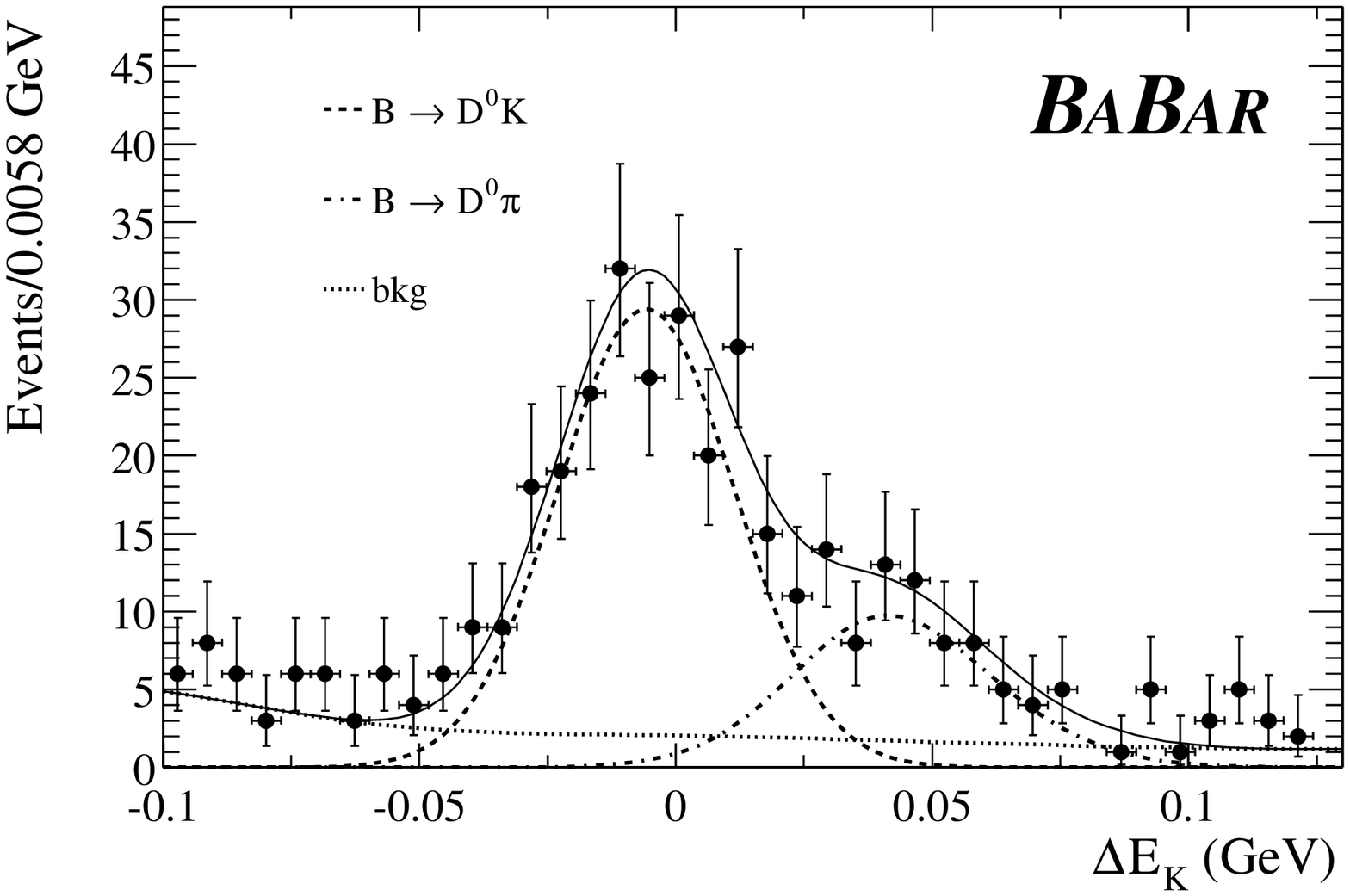}
\includegraphics[width=8cm]{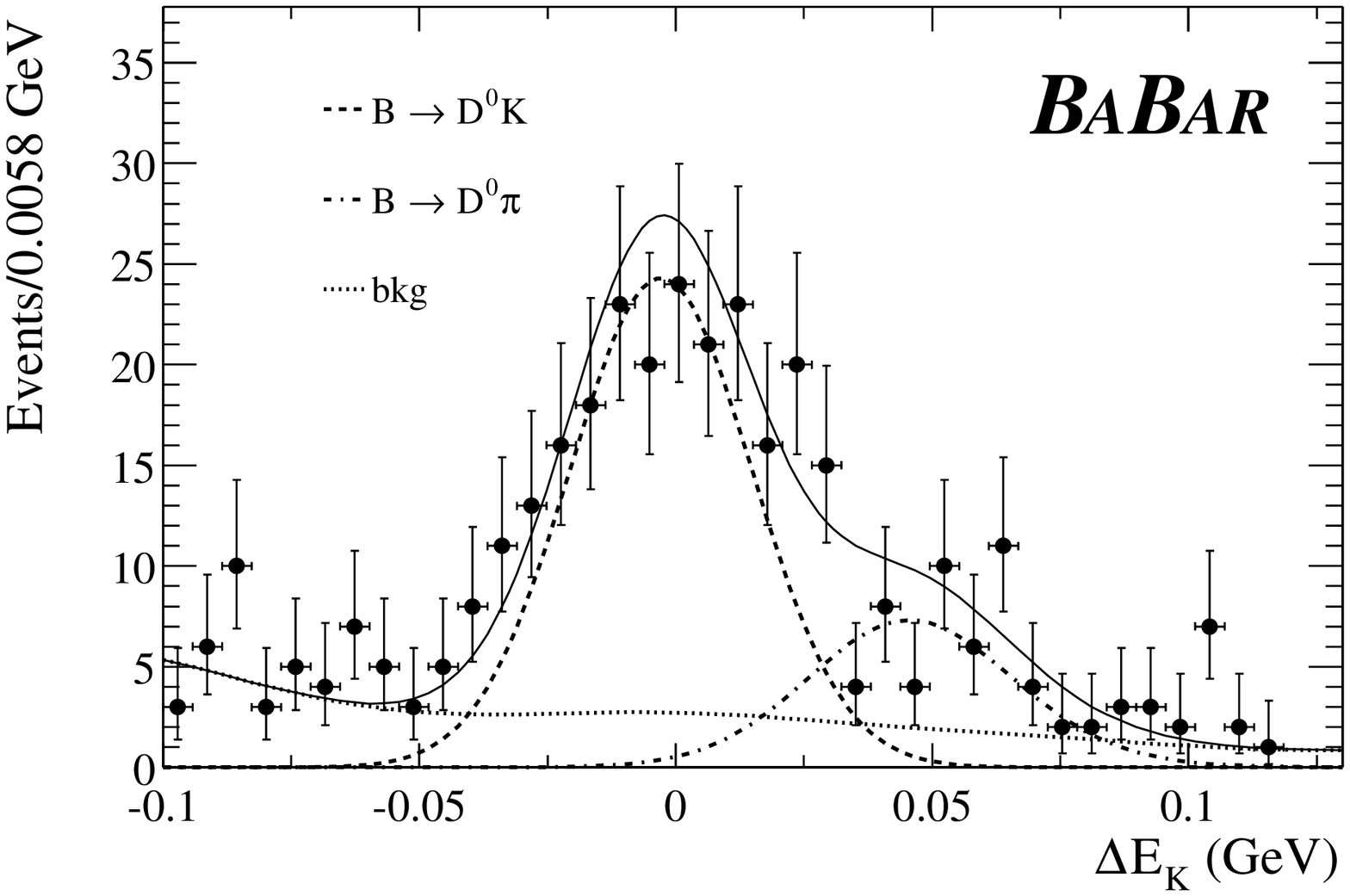}\includegraphics[width=8cm]{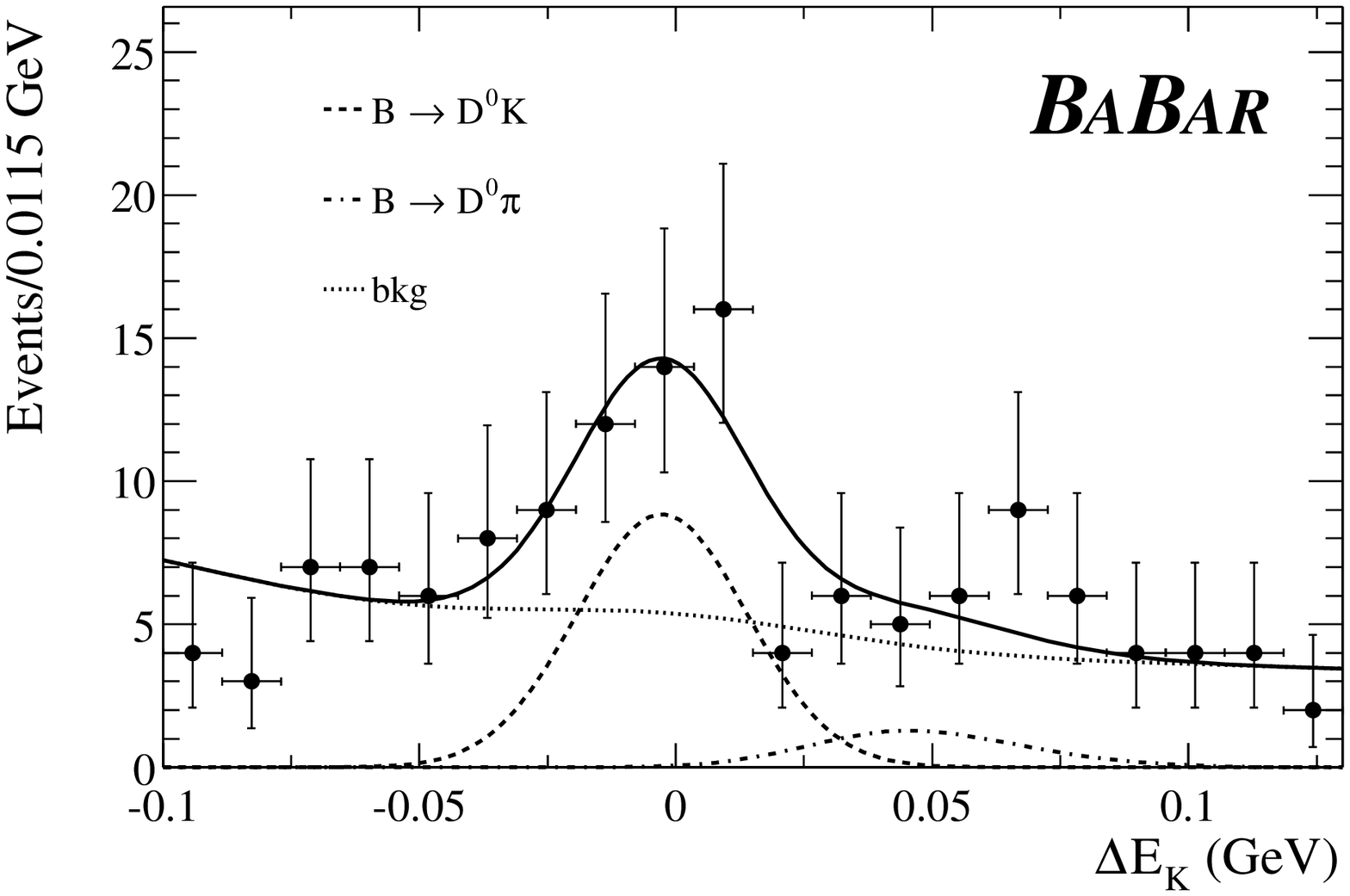}
\caption{\deltaemk\ distribution of the \btodh\ candidates selected in
the data sample by requiring that the prompt track $h$ be identified as a kaon
and $|\mes-\langle \mes \rangle|<3\sigma$. Top left: \dotokp; top right: \dotokppp; bottom left: \dotokppo; bottom right: \dotokk. The peak around zero
from \btodk\ candidates is visible.}
\label{fig:signal_fit}
\end{center}
\end{figure}

In order to increase the relative fraction of signal \btodk\ events for
illustration only, candidates are selected with the further requirements
that the prompt track be identified as a kaon and $|\mes - \langle
\mes \rangle|<3\sigma$. The \deltaemk\ distributions of the selected
candidates are shown in Figure~\ref{fig:signal_fit} for all modes.
The peak around zero from \btodk\ candidates is visible, but the
efficiency for the signal is decreased by 13\% with respect to the
standard selection outlined in Secs.~\ref{sec:Analysis} and~\ref{sec:bkg}.

\begin{table}[h]
\caption{Results from the maximum-likelihood fit. For the \dotokk
mode we quote the results for the fits performed on the whole sample
and on the $B^+$ and $B^-$ subsamples.
%The candidates for
%the non-\CP\ modes have been selected on the 2000-2001 data sample;
%the candidates for the \dotokk\ mode have been selected on the full
%2000-2002 data sample. 
}
\label{tab:fitresults}
\begin{center}
\begin{tabular}{l|c|c|c}
\hline
\hline
$D^0$ decay mode & $N($\btodp$)$ & $N($\btodk$)$ &$\int \mathcal{L}\,dt\
(\textrm{fb}^{-1})$\\
\hline
$K^-\pi^+$ & $4440\pm 69$ & $360 \pm 21$ &56.0\\
$K^-\pi^+\pi^+\pi^-$ & $2914 \pm 56$ & $242 \pm 18$& 56.0\\
$K^-\pi^+\pi^0$ & $2650\pm 56$ & $208 \pm 18$& 56.0\\ 
\hline
$K^-K^+$  & $508 \pm 24$ & $37 \pm 8$& 74.8\\
$K^-K^+$ [$B^+$] & $254 \pm 17$ & $15 \pm 6$& 74.8\\
$K^-K^+$ [$B^-$] & $254 \pm 17$ & $22 \pm 6$&74.8\\
\hline
\hline
\end{tabular}
\end{center}

\end{table}


\section{Systematic studies}
\label{sec:Systematics}
Systematic uncertainties on the ratio $R$ and on the \CP\ asymmetry
$A_{\CP}$ arise primarily from 
uncertainties on $n_i$  due to imperfect knowledge of the PDF
shapes. In this section we describe how the different contributions to
these uncertainties have been evaluated; the results are summarized in
Tables~\ref{tab:syst_ratio} and~\ref{tab:syst_asym}.

In the case of the \mes\ and \deltaemk\ PDFs for the different
signal and continuum background contributions, we vary the PDF
parameters --- which have been determined through a fit to \deltaemk\ and
\mes\ distributions in data or Monte Carlo events --- by one
statistical error.

The systematic uncertainty from the \BB\ background must be treated as a
special case, because the parameterization is performed through the
{\em Kernel Estimation\/} and not through an analytical
function. 
To study the systematic
uncertainty associated with the uncertainty on the shape of the \BB\
background we generate 500 different samples of \BB\ events and for
each one the corresponding PDF is used in the maximum-likelihood
fit. The width of the distribution 
of the difference between the new yields and the original yields is used
as an estimate of the systematic uncertainty.
Uncertainties arising from the imperfect
knowledge on the branching fractions of the different channels that
contribute to the \BB\ background are also taken into account (this is
important for the \dotokk\ mode, where one of the main sources of
uncertainty is the expected number of $B\ra KKK$ background events).

The parameterization of the particle identification PDF is performed by fitting
with a Gaussian function the background-subtracted distribution of the
difference between the reconstructed and expected Cherenkov angle of
the charged tracks from \Dz\ decays in the  $D^{*+}\to\Dz\pip$
($\Dz\to \Km\pip$) control sample, in bins of momentum and polar angle.
Therefore, the significant parameters for each momentum-polar angle
bin are the mean and the width of the fitted Gaussian.
We estimate the systematic error associated with the particle
identification PDF by
varying by one statistical error the Gaussian parameters of each bin, while
all the others are kept fixed at their central value.

The various sources of systematic errors are assumed
to be uncorrelated. The total systematic error is obtained
by summing in quadrature the individual contributions.

Systematic uncertainties on the charge asymmetries are added in
quadrature with the limit on intrinsic charge bias in the detector
(0.05), measured on data for the processes \btodp\ [\dotokp], \btodk\
[\dotokp] and \btodp\ [\dotokk], where the \CP\ asymmetry is expected
to be zero.

\begin{table}[!htb]
\caption{Systematic uncertainties on the ratio $R$ of the branching fractions
of \btodk\ and \btodp.}
\label{tab:syst_ratio}
\begin{center}
\setlength{\extrarowheight}{3pt}
\begin{tabular}{c|c}
\hline
\hline
Parameter&Uncertainty on R (\%)\\
\hline
$\Delta E_K (B^-\ra D^0h^-) $&$\pm 0.13$ \\
$m_{\rm ES} (B^-\ra D^0h^-) $&$\pm 0.01$ \\
$\qqbar$ background \deltaemk&$\pm 0.01$ \\
$\qqbar$ background \mes&$\pm 0.01$ \\
$\BB$ background \deltaemk\ {\em vs} \mes& $\pm 0.04$ \\
Particle identification&$\pm 0.16$ \\
Background composition& $\pm 0.03$ \\
\hline
\textbf{Total}&$\pm 0.20$ \\
\hline
\hline
\end{tabular}
\end{center}
\end{table}

\begin{table}[!htb]
\caption{Systematic uncertainties on the \CP\ asymmetry in the decay \btodk, with \dotokk.}
\label{tab:syst_asym}
\begin{center}
\setlength{\extrarowheight}{3pt}
\begin{tabular}{c|c}
\hline
\hline
Parameter&Uncertainty on $A_{\CP}\ (\%)$\\
\hline
$\Delta E_K (B^-\ra D^0h^-) $&$^{+0.5}_{-0.0}$ \\
$m_{\rm ES} (B^-\ra D^0h^-) $&$^{+0.5}_{-0.1}$ \\
$\qqbar$ background \deltaemk&$^{+0.3}_{-0.0}$ \\
$\qqbar$ background \mes&$^{+0.1}_{-0.7}$ \\
$\BB$ background \deltaemk\ {\em vs} \mes& $\pm 0.3$ \\
Particle identification&$^{+4.1}_{-0.9}$ \\
Background composition& $^{+6.2}_{-4.4}$ \\
Detector charge asymmetry & $\pm 5.0$\\
\hline
\textbf{Total}&$^{+9.0}_{-6.8}$ \\
\hline
\hline
\end{tabular}
\end{center}
\end{table}

\section{Results}
\label{sec:Physics}

\subsection{Measurement of the ratio $\BR(B^-\ra D^0 K^-)/\BR(B^-\ra D^0 \pi^-)$}\label{sec:br_ratio}
The ratio of the branching fractions of the decays \btodp\ and \btodk\
is calculated separately for the three non-\CP\ $D^0$ decay channels
and for the \dotokk\ mode. The ratio is computed from the number of
\btodk\ and \btodp\ mesons estimated with the maximum-likelihood fit
and listed in Table~\ref{tab:fitresults}. 
The resulting ratios are scaled
by a correction factor that account for small differences in the
efficiency between \btodk\ and \btodp\ selection, estimated with 
signal Monte Carlo samples. 
The results for the ratio of the branching fraction of the decay
\btodp\ and \btodk\ with \Dz\ decaying to the non-\CP\ modes, measured
with 56.0~\invfb, are
listed in Table~\ref{tab:final_ratio}; 
they represent three independent measurements of the same quantity, and
they are statistically consistent.  
The weighted mean of the three measurements gives the result for the
non-\CP\ modes:
\begin{equation}
R=(8.31\pm 0.35\pm 0.20)\%\ . 
\end{equation}
The resulting ratio of the branching fractions of the \dotokk\ mode,
measured with 74.8 fb$^{-1}$, is also reported in
Table~\ref{tab:final_ratio} and is found to be:
\begin{equation}
R_{\CP}=(7.4\pm 1.7 \pm 0.6)\%\ . 
\end{equation}
In the evaluation of the statistical error on $R$ the correlation
between the numbers of \btodk\ and \btodp\ events obtained from  
the fit has been taken into account.

\begin{table}[h]
\caption{Measured ratio \BR(\btodk)/\BR(\btodp) for
different \Dz\ decay modes. 
%The values for
%the non-\CP\ modes have been obtained on the 2000-2001 data sample;
%the value for the \dotokk\ mode has been measured on the full
%2000-2002 data sample. 
}
\label{tab:final_ratio}
\begin{center}
\begin{tabular}{c|c|c}
\hline
\hline
\btodh\ decay mode &\textbf{ratio} &$\int\mathcal{L}\,dt \ (\textrm{fb}^{-1})$\\
\hline
\hline
\dotokp& $\left(8.4\pm 0.5 \pm 0.2  \right)\% $ &56.0 \\
\dotokppp&$ \left(8.7\pm 0.7 \pm 0.2 \right)\% $  &56.0\\
\dotokppo& $\left(7.7\pm 0.7 \pm 0.3 \right)\% $ &56.0 \\
\hline
weighted mean& $\left(8.31\pm 0.35\pm 0.20\right)\%$ &56.0 \\
\hline
\hline
\dotokk & $\left(7.4\pm 1.7 \pm 0.6\right)\%$ &74.8\\
\hline
\hline
\end{tabular}
\end{center}
\end{table}

\subsection{Measurement of the direct \CP\ asymmetry}\label{sec:direct_asym}
The direct \CP\ asymmetry 
for the \btodk, \dotokk\ decay is calculated from the measured
yields of positive and negative decays reported in
Table~\ref{tab:fitresults}. 
The resulting asymmetry, measured with 74.8~\invfb, is
\begin{equation}
A_{\CP}= 0.17\pm 0.23\, ^{+0.09}_{-0.07} \ .
\end{equation}
We checked for detector charge bias by measuring the asymmetries for
the processes \btodp\ [\dotokp], \btodk\ [\dotokp] and \btodp\ [\dotokk], where
the \CP\ asymmetry is expected to be zero. The measured values, $(-2.4\pm1.3)\%$, $(-0.6\pm5.0)\%$ and $(0.0\pm4.7)\%$, are consistent with zero.

\section{Summary}
\label{sec:Summary}
The \btodk\ decays with $D^0$ selected in the channels \dotokp, \dotokppp, \dotokppo\ have been reconstructed on a data sample of 56.0 \invfb.
The ratio $R$ of the branching fractions $\BR(B^-\ra D^0 K^-)$ and
$\BR(B^-\ra D^0 \pi^-)$ has been measured to be
\begin{eqnarray}
R\equiv\frac{\BR(B^-\ra D^0 K^-)}{\BR(B^-\ra D^0 \pi^-)}=(8.31\pm 0.35\pm 0.20)\%\ . \nonumber
\end{eqnarray}
The \btodk\ decays with $D^0$ decaying to the \CP-even eigenstate
\dotokk\ have been reconstructed on a data sample of 74.8 \invfb. The
yield has been measured separately for positive and negative $B$
mesons. The total yield is
\begin{eqnarray}
&&N_\pm\equiv N_{B^\pm \ra D^0_{\CP}K^\pm}=36.8\pm 8.4 \pm 4.0\ . \nonumber
\end{eqnarray}
The ratio $R_{\CP}$ of the branching fractions $\BR(B^\pm\ra D^0_{\CP} K^\pm)$ and $\BR(B^\pm\ra D^0_{\CP} \pi^\pm)$ has been measured:
\begin{eqnarray}
R_{\CP}\equiv\frac{\BR(B^-\ra D^0_{\CP} K^-)+\BR(B^+\ra D^0_{\CP} K^+)}{\BR(B^-\ra D^0_{\CP} \pi^-)+\BR(B^+\ra D^0_{\CP} \pi^+)}=(7.4\pm 1.7\pm 0.6)\%\ . \nonumber
\end{eqnarray}
The direct \CP\ asymmetry has been measured:
\begin{eqnarray}
A_{\CP}\equiv\frac{\BR(B^-\ra D^0_{\CP} K^-)-\BR(B^+\ra D^0_{\CP} K^+)}{\BR(B^-\ra D^0_{\CP} K^-)+\BR(B^+\ra D^0_{\CP} K^+)} = 0.17\pm 0.23\, ^{+0.09}_{-0.07}\ . \nonumber
\end{eqnarray}

\section{Acknowledgments}
\label{sec:Acknowledgments}

We are grateful for the 
extraordinary contributions of our \pep2\ colleagues in
achieving the excellent luminosity and machine conditions
that have made this work possible.
The success of this project also relies critically on the 
expertise and dedication of the computing organizations that 
support \babar.
The collaborating institutions wish to thank 
SLAC for its support and the kind hospitality extended to them. 
This work is supported by the
US Department of Energy
and National Science Foundation, the
Natural Sciences and Engineering Research Council (Canada),
Institute of High Energy Physics (China), the
Commissariat \`a l'Energie Atomique and
Institut National de Physique Nucl\'eaire et de Physique des Particules
(France), the
Bundesministerium f\"ur Bildung und Forschung and
Deutsche Forschungsgemeinschaft
(Germany), the
Istituto Nazionale di Fisica Nucleare (Italy),
the Research Council of Norway, the
Ministry of Science and Technology of the Russian Federation, and the
Particle Physics and Astronomy Research Council (United Kingdom). 
Individuals have received support from 
the A. P. Sloan Foundation, 
the Research Corporation,
and the Alexander von Humboldt Foundation.


\begin{thebibliography}{99}


\bibitem{gronau1991} M.~Gronau and D.~Wyler, \jpl{B265}, 172 (1991); M.~Gronau
and D.~London, \jpl{B253} 483 (1991).

\bibitem{atwood1997} D.~Atwood, I.~Dunietz and A.~Soni, \jprl{78}, 3257 (1997).

\bibitem{cleo_btdk} M.~Athanas $et\ al.$, \jprl{80}, 5493 (1998).

\bibitem{belle_btdk} The Belle Collaboration, K.~Abe {\em et al.},
\jprl{87}, 111801 (2001). 

\bibitem{belle_bcptdk} BELLE Preprint 2002-19, submitted to \jprlBase

\bibitem{ref:babar}
The \babar\ Collaboration, A.\ Palano {\em et al.},
Nucl.\ Instrum.\ Methods {\bf A479}, 1 (2002).

\bibitem{ref:PDG} Particle Data Group, D.~E.~Groom {\em et al.}, Eur.\
Phys.\ J.\ C~{\bf 15}, 1 (2000).

\bibitem{ref:fox_wol} G.~C.~Fox and S.~Wolfram, \jprl{41}, 1581 (1978).

\bibitem{ref:belle_btohhh} The Belle Collaboration, K.~Abe {\em et
al.}, hep-ex/0201007, submitted to \jprBase\ D.

\bibitem{ref:kernel1} K.~S.~Cranmer, ``Kernel Estimation for
Parametrization of Discriminant Variable Distributions'', ALEPH 99-144.

\bibitem{ref:argus} ARGUS Collaboration, H. Albrecht {\em et al.},
\zp{C48}, 543 (1990).

\end{thebibliography}
\end{document}